\documentclass[twocolumn,secnumarabic,amssymb,amsmath,showpacs,nobibnotes, aps, prl]{revtex4-1}
\usepackage{graphicx}
\usepackage{amsmath}
\begin{document}
\title{Novel hyperbolic metamaterials based on multilayer graphene structures}
\author{Ivan V.~Iorsh$^{1,2}$}
\author{Ivan S.~Mukhin$^{1,3}$}%
\author{Ilya V. Shadrivov$^{4}$}
\author{Pavel A.~Belov$^{1,5}$}%
\author{Yuri S. Kivshar$^{1,4}$}
\affiliation{$^1$National Research University of Information Technologies, Mechanics and Optics (ITMO), St.~Petersburg 197101, Russia}
\affiliation{%
$^2$ Department of Physics, Durham University, DH1 3LE Durham, UK }
\affiliation{$^3$St.~Petersburg Academic University, Nanotechnology Research and Education Center, St.~Petersburg 194021, Russia}

\affiliation{$^4$Nonlinear Physics Center, Research School of Physics and Engineering, Australian National University, Canberra ACT 0200, Australia}

\affiliation{$^5$Queen Mary University of London, London E1 4NS, UK}


\begin{abstract}
We suggest a new class of hyperbolic metamaterials for THz frequencies based on multilayer graphene structures. We calculate the dielectric permittivity tensor of the effective nonlocal medium with a periodic stack of graphene layers and demonstrate that tuning from elliptic to hyperbolic dispersion can be achieved with an external gate voltage. We reveal that such graphene structures can demonstrate a giant Purcell effect that can be used for boosting the THz emission in semiconductor devices. Tunability of these structures can be enhanced further with an external magnetic field which leads to the unconventional hybridization of the TE and TM polarized waves.
\end{abstract}

\maketitle

A hyperbolic medium is a special class of indefinite media~\cite{smith2003} described by the diagonal permittivity tensor with the principal components being of the opposite signs which results in a hyperbolic shape of the isofrequency contours~\cite{Xie2009,narimanov2010}. Such media have a number of unique properties including negative refraction~\cite{smith2003,nrefraction} and subwavelength imaging~\cite{hyperlens}. One of the possible realizations of hyperbolic media is a periodic metal-dielectric nanostructured metamaterial where the hyperbolic nature of the isofrequency curves appears due to the excitation of the near-field plasmon Bloch waves~\cite{Vinogradov,PLA}. Hyperbolic metamaterials have been realized for optical, infrared, and microwave frequency ranges. Realization of the THz hyperbolic media could allow to boost otherwise slow THz radiative transitions in semiconductor devices which would lead to the development of a new class of THz sources.

Graphene, a two-dimensional lattice of carbon atoms, exhibits a wide range of unique properties~\cite{Rev1,Rev2,Rev3}. Surface plasmons excited in individual graphene sheets have been extensively studied, both theoretically~\cite{theor2,theor1,theor3,theor4,theor5,Engheta,Newmode} and experimentally~\cite{Koppens_exp,Basovexp}.

\begin{figure}[!h]
\centerline{\includegraphics[width = 0.9\columnwidth]{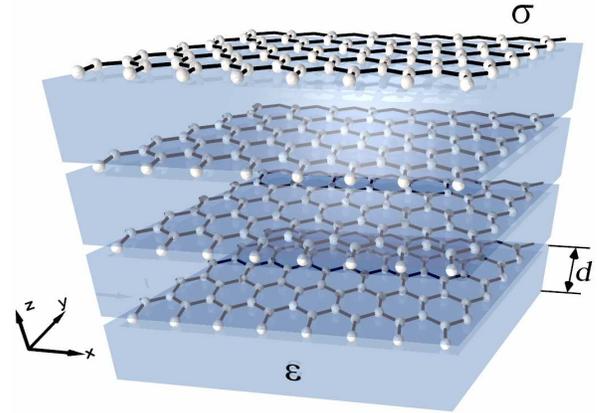}}
\caption{(Color online) Geometry of the effective hyperbolic metamaterial composed of a periodic structure
of graphene layers with conductivity $\sigma$ separated by dielectric slabs with the thickness $d$ and dielectric constant $\epsilon$.}
\label{fig1_graph}
\end{figure}

In this Letter, we suggest a novel class of hyperbolic metamaterials where individual graphene sheets are separated by host dielectric slabs, as shown schematically in Fig.~\ref{fig1_graph}. It is easy to notice an analogy between a graphene sheet placed inside a dielectric  medium and a thin metal waveguide imbedded into a dielectric matrix, which also supports localized surface plasmon polaritons. Assuming this analogy, we may expect that a periodic lattice of the graphene sheets may behave like an effective hyperbolic medium due to the coupling between the surface plasmons localized at the individual graphene sheets~\cite{new_PRL}. Importantly, surface plasmons in graphene have low losses and strong localization in the THz region. Indeed, as we demonstrate below, a periodic structure of graphene layers creates {\em a novel type of metamaterial} with strong nonlocal response and hyperbolic properties of its dispersion curves for TM-polarized waves in the THz frequency range and superior characteristics such as a giant Purcell effect and tunability by a gate voltage or magnetic field.

It is important to mention that the periodic layered structure shown in Fig.~\ref{fig1_graph} resembles a natural graphite which is known to exhibit many interesting properties in the ultraviolet frequency range~\cite{graphite}. However, in the case of graphite, the $\pi$-orbitals of the carbon atoms in the individual graphene planes strongly overlap which results in the modification of the electron band structure and large nonradiative losses. In the metamaterials we suggest here the electronic structure of the graphene sheets remains unperturbed. 

We consider the structure shown in Fig.~\ref{fig1_graph} composed of graphene layers separated by dielectric slabs. Recently a graphene based hyperlens has been proposed in \cite{Lavrinenko}. However, the structure considered in this paper resembles the wire medium metamaterial and our structure is similar rather to the mutlilayered meta-dielectric metamaterials. First, we employ the transfer matrix approach and analyze the dispersion properties of electromagnetic waves for both polarizations. We describe a graphene sheet by macroscopic parameters as a conducting surface defined by the frequency dependent conductivity $\sigma(\omega)$~\cite{Geimcond}. Electromagnetic boundary conditions at the graphene plane can be written as a matrix $\hat{M}$ connecting the tangential components of the electromagnetic field from both sides,
\begin{align}
\hat{M}_{\rm TE,TM}=\begin{pmatrix} 1 & 0 \\ - 4\pi\sigma/c & 1 \end{pmatrix},
\end{align}
so that the transfers matrix for the structure period can be defined as $\hat{T}=\hat{S}\hat{M}\hat{S}$, where the matrix $\hat{S}$ is the transfer matrix of a dielectric layer. Then, we obtain the dispersion relation for the Bloch waves in the metamaterial structure in the form $\cos(Kd)=(\hat{T}_{11}+\hat{T}_{22})/2$, i.e.
\begin{align}
TE:\quad \cos(Kd)=\cos(k_z d)-\frac{2i \pi \sigma k_0}{k_z c }\sin(k_z d), \label{bloch1} \\
TM:\quad \cos(Kd)=\cos(k_z d)-\frac{2i \pi \sigma k_z}{k_0 c \varepsilon }\sin(k_z d), \label{bloch2}
\end{align}
where $k_z=(\varepsilon k_0^2-\beta^2)^{1/2}$, $\beta$ is the wavevector in the plane, $K$ is the Bloch wavenumber, and $k_0=\omega/c$.

\begin{figure}[!h]
\centerline{\includegraphics[width = 1.0\columnwidth]{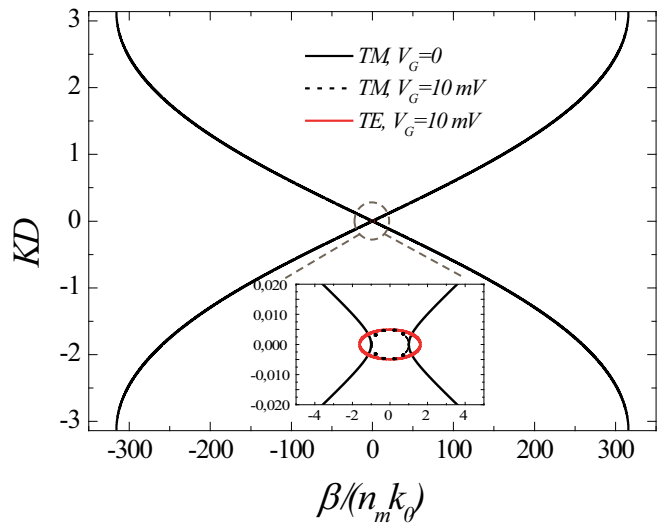}}
\caption{(Color online) Isofrequency contours of the graphene metamaterial for the different values of the gate voltage and a fixed frequency of $4$ meV.  }
\label{fig2}
\end{figure}

As is clear from Eqs.\eqref{bloch1} and \eqref{bloch2}, the dispersion properties of the structure are defined by the graphene conductivity $\sigma(\omega)$ which depends on the chemical potential $\mu$. Thus, changing $\mu$ with the external gate voltage we can tune the isofrequency contours of the metamaterial. 
For example, for the chemical potential the isofrequency curves of TM polarized waves are elliptic whereas
they become hyperbolic for the gate voltage to $10$ mV (see Fig.~\ref{fig2}). The transition from the elliptical to hyperbolic regime has been studied recently in metamaterials \cite{topolog}, and the graphene-based metamaterials offer a simple way for the realization of such transitions.

As has been established recently~\cite{Orlov}, dispersion properties of nanostructured metal-dielectric metamaterials may demonstrate strong optical nonlocality due to excitation of surface plasmon polaritons, and these effects are more pronounced when the thicknesses of metal and dielectric layers are dissimilar.
Accordingly, we expect graphene metamaterials to exhibit strong nonlocal properties having one of the layers vanishingly thin.

To derive the effective parameters characterizing the graphene metamaterial, we employ the nonlocal homogenization procedure~\cite{Chebykin}. Within this approach, we assume that the structure is excited with a harmonic external current $J\sim \exp[i(\beta x+K z-\omega t)]$, and we calculate the electrical displacement averaged over the period of the structure.
\begin{figure}[!h]
\centerline{\includegraphics[width = 1.0\columnwidth]{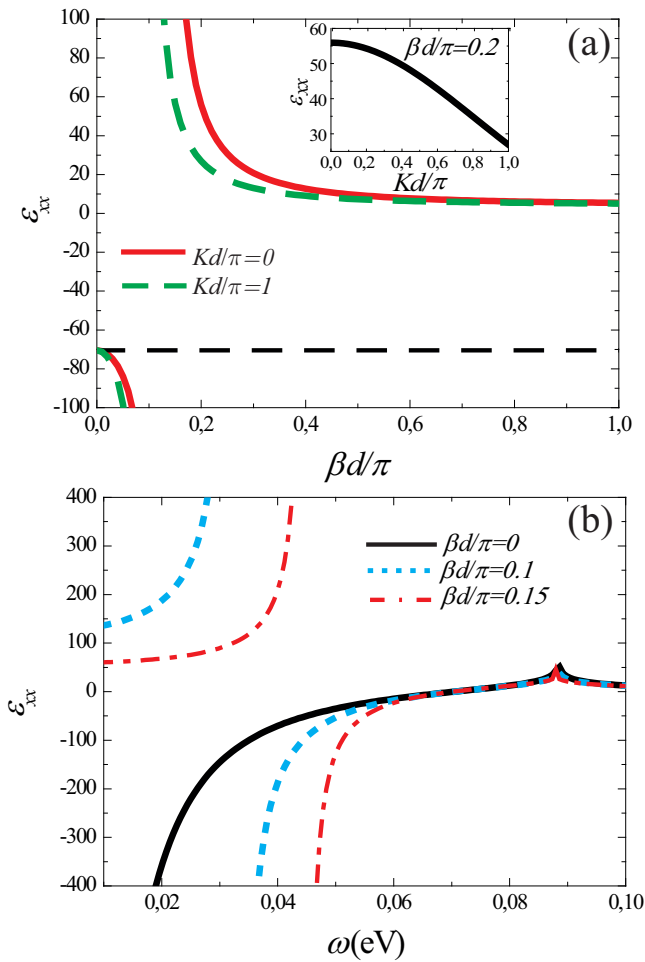}}
\caption{(Color online) (a) Dependence of  the component $\varepsilon_{xx}$ on the in-plane wavevector $\beta$. Frequency is $40$ meV; local approximation result is shown with a straight dashed line. Insert shows the dependence on $K$. (b) Frequency dependence of the component $\varepsilon_{xx}$ for different values of $\beta$.
}
\label{figepsilon}
\end{figure}
After lengthy calculation employing the techniques from Ref.~\cite{Chebykin},
we obtain the nonlocal dielectric permittivity tensor $\hat{\varepsilon}_{\rm nloc}$,
\begin{align}
\hat{\varepsilon}_{\rm nloc}=\begin{pmatrix} \varepsilon_{xx} & 0 & 0 \\ 0 & \varepsilon_{yy} & 0 \\ 0 & 0 & \varepsilon_{zz} \end{pmatrix},
\end{align}
with all nondiagonal elements vanishing and the remaining three components
\begin{align}
\varepsilon_{xx} =\varepsilon - 2\alpha\left\{1-2\alpha \frac{k_z^2}{k_0^2} f(\beta,K) \right\}^{-1}, \\
\varepsilon_{yy} =\varepsilon - 2\alpha\biggr[1-2\alpha f(\beta,K)\biggr]^{-1}, \\
\varepsilon_{zz} = \varepsilon, \label{epsilons}
\end{align}
where
$\alpha=2\pi\mathrm{Im}(\sigma)/(k_0dc)$, and the function $f(\beta,K)$ has the form,
\begin{align}
f(\beta,K)=\frac{k_0^2d \sin(k_zd)}{2 k_z [\cos(Kd)-\cos(k_z d)]}-\frac{k_0^2}{(k_z^2-K^2)},
\end{align}

For small values of $\beta$ and $K$, the function $f(\beta,K)$ vanishes, and the components of the effective dielectric permittivity tensor become wave independent as in the case of quasistatic approximation and local media, $\varepsilon_{xx}=\varepsilon_{yy}=\varepsilon-2\alpha$.

Figure~\ref{figepsilon}(a)  shows the dependence of the dielectric tensor component $\varepsilon_{xx}=\varepsilon-2\alpha$.  on the in-plane wavevector $\beta$ for a fixed frequency of $40$ meV, whereas the prediction of the local approach is shown with a straight dashed line. The component $\varepsilon_{yy}$ deviates slightly from its local value $\epsilon-2\alpha$, and it is not shown in the figure.

We observe that the nonlocal dielectric permittivity becomes negative for small values of $\beta d$. Because the component $\varepsilon_{zz}$ is always positive, this results in a hyperbolic shape of the isofrequency contours. However, for larger values of $\beta$ the local approximation becomes invalid, and the component $\varepsilon_{xx}$ changes its sign to positive, so an effective hyperbolic medium is transformed into elliptic one. This is a direct manifestation of strong nonlocal properties of the graphene metamaterials.

Figure~\ref{figepsilon}(b) shows the dependence of the component $\varepsilon_{xx}$ vs. frequency. A solid line corresponds to the case of $\beta=0$, and it coincides with the result of the local approximation. In this case, $\varepsilon_{xx}$ may take very large negative values. However, there exist again strong discrepancies between local and nonlocal theories for the case of nonzero $\beta$. We observe that in this case there is a certain frequency corresponding to the pole of expression~\eqref{epsilons} for $\varepsilon_{xx}$ below which the dielectric permittivity becomes positive. 

\begin{figure}[!h]
\centerline{\includegraphics[width = 1.0\columnwidth]{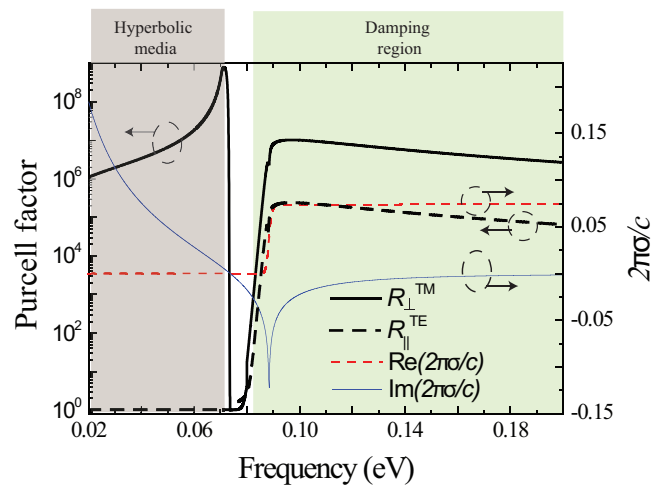}}
\caption{(Color online) Purcell factor vs frequency for both TE and TM polarizations. Temperature is $4$ K. }
\label{fig3}
\end{figure}

To illustrate some consequences of these results, we calculate the Purcell factor of a point dipole placed on top of the graphene metamaterial which characterizes the enhancement of the radiation efficiency as compared to the free space. In our structure we expect large Purcell factors for low frequencies. The expressions for the Purcell factors for both dipole orientations have the form~\cite{PLA},
\begin{align}
R_{\|}^{TE}=\frac{3}{4}\mathrm{Re}\int\limits_{0}^{\infty}\frac{\beta d\beta}{k_0\sqrt{k_0^2-\beta^2}}r_{TE}, \label{intte} \\
R_{\|}^{TM}=\frac{3}{4}\mathrm{Re}\int\limits_{0}^{\infty}\frac{\beta\sqrt{k_0^2-\beta^2} d\beta}{k_0^3}r_{TM}, \label{inttm1} \\
R_{\perp}^{TM}=\frac{3}{4}\mathrm{Re}\int\limits_{0}^{\infty}\frac{\beta^3 d\beta}{k_0^3 \sqrt{k_0^2-\beta^2}}(-r_{TM}), \label{inttm2}
\end{align}
where signs ${\|}$ and ${\perp}$ correspond to the parallel and perpendicular orientations to the interface, respectively, and $r_{TE}, r_{TM}$ are the reflection coefficients from the semi-infinite layered structure. 
The Purcell factor is shown in Fig.~\ref{fig3}. We observe that the low frequency range is characterized by large Purcell factors for TM-polarized waves. This agrees with our observation of hyperbolic isofrequency contours for the low frequency range in TM polarization. We also notice that there are no large Purcell factors values for the TE polarization in low-loss frequency range. Moreover, for the large frequencies the large values of Purcell factors can be observed for both polarizations. This increase in Purcell factor is connected with a step-like increase of the non-radiative losses in the graphene (proportional to the real part of conductivity). Here we should mention the effect of the temperature on the spontaneous emission enhancement. The temperature defines the steepness of the step-like increase of the real part of the conductivity. In order to minimize the non-radiative losses in the hyperbolic media region we should imply the condition $\mu/(k_B T)\ll 1$. In the THz frequency range, this condition corresponds to the temperatures of about $10$ Kelvin.

We have also obtained analytical expressions for the $R_{\perp}^{TM}$ in two limiting cases. When the conductivity of graphene is small, the Purcell factor takes the form,
\begin{align}
R_{\|}^{TM}\approx \frac{3\pi\varepsilon^2}{2} \left(\frac{c}{2\pi |\mathrm{Im}(\sigma)|} \right)^3 \exp\left(-\frac{\varepsilon c k_0 d}{2\pi |\mathrm{Im}(\sigma)|} \right), \label{anal2}
\end{align}
We notice that the Purcell factor is approaching zero exponentially when conductivity vanishes. 
When the conductivity is large, the expression for $R_{\perp}^{TM}$ reads
\begin{align}
R_{\|}^{TM}\approx \frac{3\pi\varepsilon}{8(k_0d)}\left(\frac{c}{2\pi|\mathrm{Im}(\sigma)|}\right)^2,
\label{anal1}
\end{align}
so that in the limit of infinite conductivity the Purcell factor vanishes. 

Thus, the Purcell factor vanishes very large and very small conductivities. Thus, we should observe some maximum at the intermediate case. We have revealed that the   maximum value of the Purcell factors is achieved for $\alpha=\varepsilon/2$. We notice that  $\alpha=\varepsilon/2$ corresponds to {\em epsilon-near-zero metamaterial}. In this regime, the media is characterized by almost plane isofrequency contours and all evanescent modes can be excited.

In the low frequency region the losses are small for sufficiently low temperatures. Thus, no contribution to the Purcell factor in the low frequency region is due to the losses. From the other hand, in our calculations we did not account for nonlocal effects in graphene conductivity which should become significant for the in-plane vectors of the order of the Fermi wavevector: $k_F=\mu/v_F$, where $v_F$ is the Fermi velocity of graphene electrons, and $\mu$ is the Fermi energy. We have checked how the nonlocality would modify the obtained results by performing integration in Eqs.~\eqref{intte},\eqref{inttm1}, and \eqref{inttm2} to $k_F$. We fix the frequency to $44$ meV and change the period of the structure. We find that for small periods the effective model sufficiently overestimates the values of the Purcell factor. However, when the period is increased the values of the Purcell factors for both models are almost equivalent. We have also derived an analytical condition of good agreement between local model and the model which performs integration only over a finite range of wavevectors: $\coth(\mu d/(2v_F))< {2\pi|\mathrm{Im}(\sigma)|\mu}/({\omega v_F})$. A good agreement between local and nonlocal models allows to expect strong THz radiation enhancement in multilayered graphene metamaterials.

We also study the properties of this novel metamaterial in the presence of a strong constant magnetic field. In this case, the surface conductivity of graphene is described by a $2\times 2$ tensor with nondiagonal component equal to the Hall graphene conductivity $\sigma_H$. Furthermore, in the presence of strong magnetic field the longitudinal and Hall parts of the conductivities are governed by the electron transitions between the discrete Landau levels~\cite{Gusynin}. In addition, for the case of a single graphene sheet the magnetic field couples the TE and TM eigenmodes resulting in the emergence of new magneto-plasmonic eigenmodes~\cite{PRBGraphene}.

For the layered structure, we observe the similar coupling between the TE and TM modes. Indeed, the Bloch vectors of the system obey the following equation,
\begin{align}
&\cos(K_{1,2}d)=\frac{\mathcal{A}+\mathcal{B}}{2}\pm \sqrt{\frac{(\mathcal{A}-\mathcal{B})^2}{4}-\frac{\pi^2\sigma_H^2\sin^2(k_z d)}{\varepsilon}}, \nonumber \\
&\mathcal{A}=\cos(k_z d)-\frac{2i\pi\sigma}{c}\frac{k_0}{k_z}\sin(k_z d), \nonumber \\
&\mathcal{B}=\cos(k_z d)-\frac{2i\pi\sigma}{c}\frac{k_z}{k_0\varepsilon}\sin(k_z d).
\label{coupled}
\end{align}
When the coupling term $\sim \sigma_H$ vanishes, Eq.~\eqref{coupled} simplifies to Eqs.~\eqref{bloch1} and \eqref{bloch2}. In the presence of the magnetic field, the TE and TM polarized modes become coupled which results in complicated shapes of the isofrequency contours. To illustrate these results, we plot
the eigenmodes isofrequency contours in the presence of constant magnetic field of 1~T applied along the $z$ axis.

\begin{figure}[!h]
\centerline{\includegraphics[width = 0.9\columnwidth]{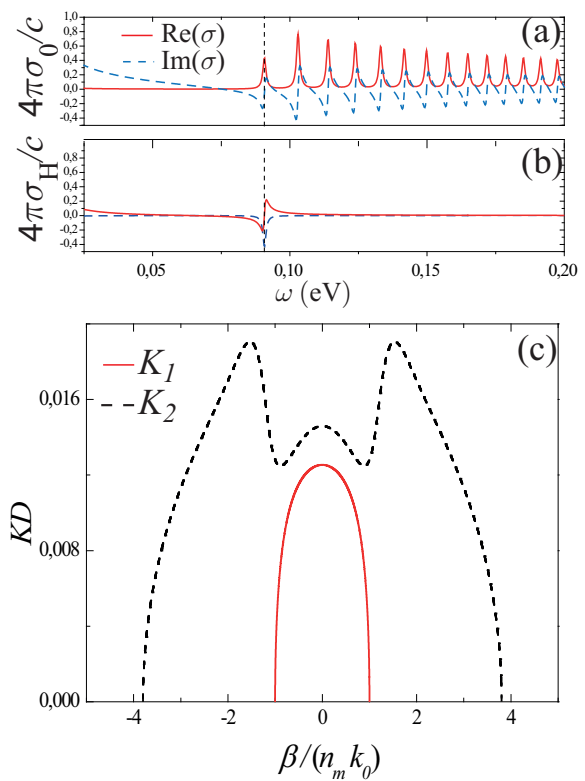}}
\caption{(Color online) (a) Frequency dependence of the Hall and longitudinal conductivities in the presence of static magnetic field of 0.1 T. Temperature is 4 K. (b) Isofrequency contours for the frequency 100 meV for the first (red solid line) and second (blue dashed line) hybrid electromagnetic modes.}
\label{fig4}
\end{figure}

 Figure~\ref{fig4}(a) shows the dispersion of the longitudinal and Hall conductivities in the vicinity of the resonance corresponding to the transition between the 0-th and 1-st Landau levels. Figure~\ref{fig4}(b) shows the isofrequency contours of the electromagnetic eigenmodes at the frequency 100 meV.
 We observe that, although the imaginary part of the longitudinal conductivity is negative and we would expect the elliptic contours for both the eigenmodes, the coupling via the Hall conductivity term changes the properties dramatically, and the isofrequency contours have a complicated shape, which is neither elliptical nor hyperbolic. The strong dependence of the isofrequency contours on the magnetic field suggests a new degree of freedom for engineering optical properties of these novel metamaterials.

Finally, we comment briefly on feasibility of the predicted effects.  It is technologically achievable to make layered structures consisting of several graphene sheets separated by dielectric layers.  For fabrication of single, double and multiple graphene layers, or two-dimensional layers of
other materials, the most popular approaches are cleavage technique~\cite{Novoselov2} as well as CVD~\cite{Petrone} and MBE~\cite{Godev} methods. There exist several methods of layer transfers, such as wet and dry transfer procedures with PMMA, PMGI or water-soluble layer~\cite{Novoselov1, Berger,Dean}. These methods remain a lot of residue at the later after transfer, but the surface can be cleaned with Ar/H2 annealing. In a number of studies~\cite{Novoselov1, Berger,Dean}, heterostructures composed of several graphene layers, boron nitride, or molybdenum flakes have been already demonstrated. Hexagonal Boron Nitride (h-BN), conventionally used as a host media, can be employed as dielectric with the permittivity 4 and higher in the THz frequency range.

In conclusion, we have demonstrated that a multilayer periodic structure composed of graphene layers can operate as a novel type of hyperbolic metamaterial in the THz frequency range, where tuning from elliptic to hyperbolic media can be achieved by applying an external gate voltage. We have predicted a giant  Purcell effect for a point dipole source placed on top of such a metamaterial being achieved through coupling of effective plasmon modes with infinite wavenumbers. We have also demonstrated a novel possibility of tuning the dispersion properties of the metamaterials with external magnetic field and the emergence of a family of novel hybrid eigenmodes in the graphene metamaterials.

This work was supported by the Ministry of Education and Science of Russian Federation, the Dynasty Foundation, and the Australian Research Council.

\end{document}